\documentclass[fleqn,10pt]{wlscirep}
\usepackage[utf8]{inputenc}
\usepackage[T1]{fontenc}
\usepackage{graphicx}
\usepackage{xr}
\title{Artificial intelligence in communication impacts language and social relationships}

\author[1,*]{Jess Hohenstein}
\author[2]{Dominic DiFranzo}
\author[1]{Rene F. Kizilcec}
\author[2]{Zhila Aghajari}
\author[3]{Hannah Mieczkowski}
\author[1]{Karen Levy}
\author[1,4]{Mor Naaman}
\author[3]{Jeff Hancock}
\author[1,*]{Malte Jung}

\affil[1]{Department of Information Science, Cornell University, Ithaca, NY 14853, USA}
\affil[2]{Department of Computer Science and Engineering, Lehigh University, Bethlehem, PA 18015, USA}
\affil[3]{Department of Communication, Stanford University, Stanford, CA 94305, USA}
\affil[4]{Cornell Tech, New York, NY 10044, USA}

\affil[*]{jch378@cornell.edu; mfj28@cornell.edu}

\begin{abstract}
Artificial intelligence (AI) is now widely used to facilitate social interaction, but its impact on social relationships and communication is not well understood. We study the social consequences of one of the most pervasive AI applications: algorithmic response suggestions ("smart replies"). Two randomized experiments (n = 1036) provide evidence that a commercially-deployed AI changes how people interact with and perceive one another in pro-social and anti-social ways. We find that using algorithmic responses increases communication efficiency, use of positive emotional language, and positive evaluations by communication partners. However, consistent with common assumptions about the negative implications of AI, people are evaluated more negatively if they are \textit{suspected} to be using algorithmic responses. Thus, even though AI can increase communication efficiency and improve interpersonal perceptions, it risks changing users' language production and continues to be viewed negatively.
\end{abstract}
\begin{document}

\flushbottom
\maketitle

\thispagestyle{empty}

\section*{Introduction}

%Furthermore, language is also an important factor in determining conversational outcomes (e.g., linguistic features are highly predictive of whether a conversation will go awry \cite{Zhang2018}). Similarly, the way that we use language also has implications for interpersonal relationships. Previous work has indicated that certain linguistic variables are indicative of constructive and destructive conflict processes (i.e., achieving goals while maintaining or enhancing one's relationship with the opponent and inflicting psychological or physical damage on one's opponent, respectively) \cite{Samson2010}. 

Communication is the basic process through which people form perceptions of others \cite{mairesse2006automatic, mairesse2006words, mairesse2007using}, build and maintain social relationships \cite{pennebaker2003psychological}, and achieve cooperative outcomes \cite{Zhang2018}.  
Applications of artificial intelligence (AI) are increasingly shaping the way that people communicate and interact with one another \cite{stanfordReport, rahwan2019machine}.
%Applications of artificial intelligence (AI) are increasingly embedded in people's daily lives \cite{stanfordReport}.
One of the most visible AI applications is AI-generated reply suggestions in text-based communication, commonly known as smart replies, which aim to help users compose messages with "just one tap" \cite{Kannan2016}. Despite the rapid deployment of AI applications in new products and contexts and people's growing concerns about the societal consequences of AI \cite{shakirov2016review}, research has predominantly focused on the technical aspects and largely ignored the potential social impacts of integrating AI-generated messages into human-to-human communication. Reports from the AI Now Institute liken this scenario to "...conducting an experiment without bothering to note the results" \cite{crawford2016ai} and have repeatedly noted the under-investment in research on the social implications of AI technologies while calling for an increase in interdisciplinary research focusing on examining these systems within human populations \cite{crawford2016ai,campolo2017ai,whittaker2018ai}.

%A growing area of AI development focuses on building systems that aim to directly interface and collaborate with people, and one of the most widely-used examples is algorithmically-suggested responses, commonly known as smart replies. 
As of last year, algorithmic responses constituted 12\% of messages sent through Gmail alone \cite{srGmail}, representing about 6.7 billion emails written by AI on our behalf each day \cite{kraus2018mashable}. Smart reply systems aim to make text production more efficient by drawing from general and user-specific text corpora to predict what a person might type and generating one or more suggested responses that a user can choose from when responding to a message \cite{Kannan2016}. Users' rapid adoption of this type of AI in interpersonal communication has been facilitated by a large body of technical research regarding various methods for generating algorithmic responses (e.g., \cite{henderson2017efficient, Kannan2016, ritter2011data}). However, the social implications of this type of AI involvement remain largely unclear. 

Given the broad integration of AI systems like these in our social lives, a growing body of work at the intersection of computer and social science is concerned with understanding how such systems may be influencing human behavior and how they are perceived (e.g., \cite{rahwan2019machine, lee2018understanding, hancock2020ai}). Initial studies have found that algorithmic responses can impact how people write \cite{arnold2020predictive}, and people believe that the mere presence of smart replies influences the way that they communicate, in part because of the linguistic skew of smart replies, which tend to express more positive motions \cite{Hohenstein2018}. However, the social implications of smart reply use remain unclear.

To examine the social consequences of using AI to help generate messages, we conducted a set of randomized controlled experiments to study how the display and use of AI-generated smart replies in real-time text-based communication affects how people interact and perceive each other. We show that a commercially-deployed AI affects various aspects of interpersonal communication. More specifically, we find that AI influences multiple dimensions of social engagement including communication efficiency, emotional tone, and interpersonal evaluations in both positive and negative ways.

\section*{Results}

\subsection*{AI is Perceived Negatively but Improves Interpersonal Perceptions}

Inspired by theories of how computer-mediated communication can affect intimacy and relationship maintenance \cite{tongWalther2011}, we hypothesized that seeing AI-generated reply suggestions could influence participants' feelings of connectedness with their conversation partner. To test the effect of AI mediation on interpersonal trait inferences and perceptions of cooperativeness, we developed a messaging application in which we can manipulate the smart replies that are displayed to users while collecting data about the conversation. 

To identify the effects and perceptions of using algorithmic responses in conversation (beyond merely being presented with them), we randomly assigned 219 pairs of participants into three different messaging conditions: 1) both participants can use smart replies (i.e., suggested responses generated using the Google Reply API \cite{smartReplyAPI}), 2) only one participant can use smart replies, or 3) neither participant can use smart replies. Participants engaged in a conversation about a policy issue while the application tracked their use of smart replies. By presenting participants with smart replies, they were encouraged to use them in conversation, which serves as our causal identification strategy for estimating the effects of smart reply use by the self and the partner. After completing the conversation, participants were given a definition of smart replies and asked to rate how often they believed that their partner had used them. They also responded to established measures of dominance and affiliation (Revised Interpersonal Adjective Scale \cite{wiggins1988psychometric}), and perceived cooperative communication (\cite{lee1997leader}). 

\begin{figure}[h]
	\centering
	\includegraphics[width=.49\linewidth]{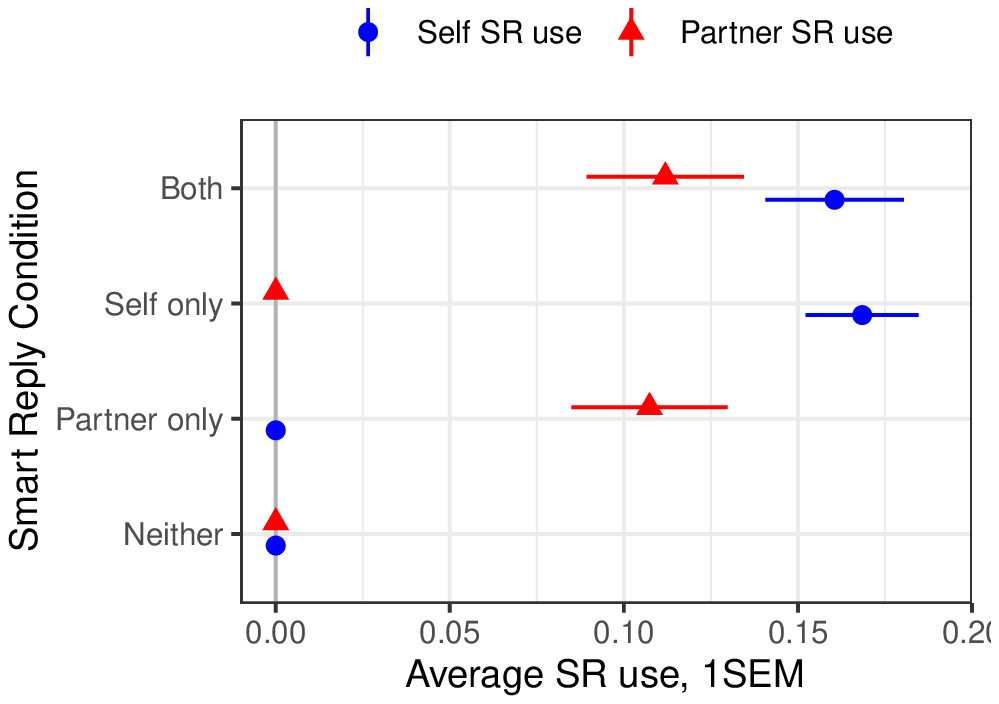}
	\includegraphics[width=.49\linewidth]{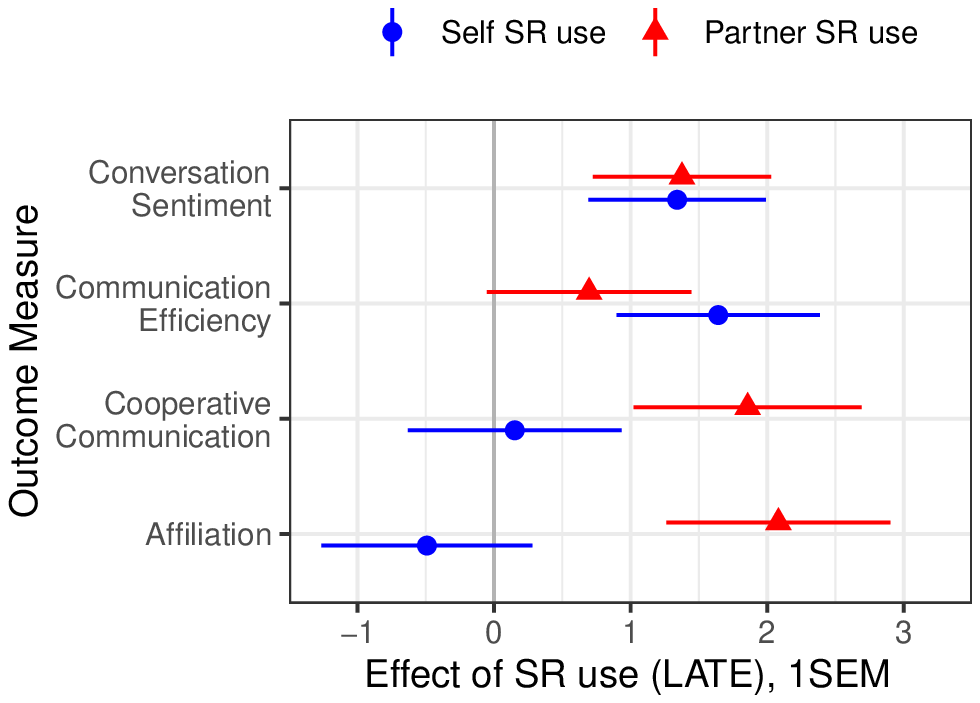}
	\caption{Results of the randomized experiment independently manipulating the availability of smart replies (SR) for each partner. Left panel: average proportion of SR use in conversation by SR condition (i.e. first-stage effect). Right panel: latent average treatment effect (LATE) estimates of how increased SR use by the self and by the partner affected conversation sentiment, communication efficiency, cooperative communication, and affiliation. Error bars represent cluster-robust standard errors.}
	\label{fig:study3}
\end{figure}

We find that the presence of algorithmic responses was a strong encouragement to use them: smart replies account for 14.3\% of sent messages on average (\textit{t}(211)=13.8, \textit{p}$<$.0001), and Figure \ref{fig:study3} (left panel) shows average smart reply use by experimental condition. Because the variation in smart reply use is experimentally and independently induced for each participant, we can use instrumental variable (IV) estimation to identify the consequences of increased smart reply use by the self and the partner (Figure \ref{fig:study3}, right panel). Using IV estimation, we find that increased use of smart replies by the self (but not the partner) led to more efficient communication in terms of the number of messages the self sent per minute (\textit{t}(198)=2.21, \textit{p}=0.0286). While smart reply use clearly improves communication efficiency, its consequences for interpersonal perceptions are more complex.

\begin{figure}[h!]
		\centering
		\includegraphics[width=.6\linewidth]{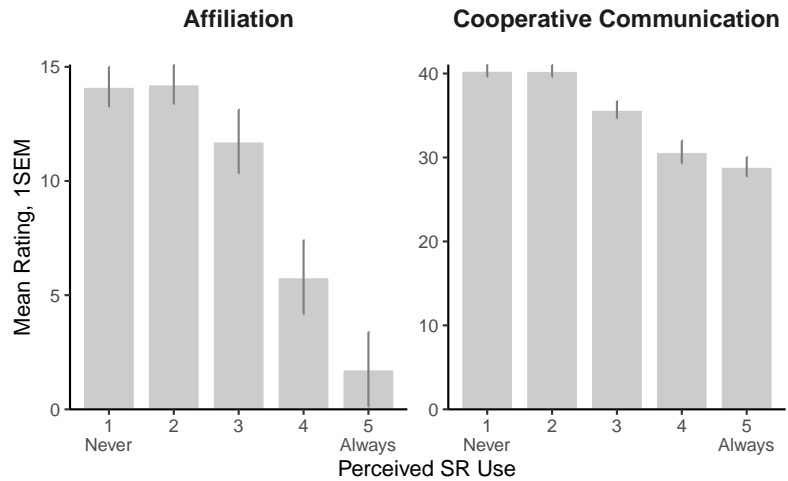}
		\caption{Average rating of the partner's affiliation and cooperative communication by the self for different levels of perceived smart reply use by the partner (\textit{N} = 361). Error bars show 1 cluster-robust standard error.}
		\label{fig:cor}
	\end{figure}

Participants are capable of recognizing their partner's use of smart replies to some degree: beliefs about how much their partner used smart replies correlated with actual use but not strongly (Pearson's \textit{r}=0.22, \textit{t}(97)=3.62, \textit{p}=0.0005). Consistent with commonly held beliefs about the negative implications of AI in social interactions \cite{Jakesch, Promberger2006}, we find strong associations between perceived smart reply use by the partner and attitudes towards them. The more participants thought their partner used smart replies, the less cooperative they rated them (\textit{t}(92)=-9.89, \textit{p}$<$0.0001) and the less affiliation they felt towards them (\textit{t}(92)=-6.90, \textit{p}$<$0.0001), as shown in Figure~\ref{fig:cor}, even after controlling for their partner's actual smart reply use. This shows correlationally that people who appear to be using smart replies in conversation ultimately pay an interpersonal toll, even if they are not actually using smart replies. However, this finding does not show causally how attitudes shift in response to actual smart reply use.

Our IV estimation strategy reveals that increased use of smart replies by the partner actually improved the self's rating of the partner's cooperation (\textit{t}(167)=2.23, \textit{p}=0.0273) and sense of affiliation towards them (\textit{t}(167)=2.54, \textit{p}=0.0120). Although perceived smart reply use is judged negatively, actual use results in more positive attitudes. Moreover, we find that conversation sentiment became more positive as a result of both the self using more smart replies (\textit{t}(198)=2.06, \textit{p}=0.0404) and the partner using more smart replies (\textit{t}(198)=2.11, \textit{p}=0.0362). This finding suggests that the effects of AI mediation on interpersonal perceptions are related to changes in language introduced by the AI system.

\subsection*{AI Sentiment Affects Emotional Content in Human Conversations}
To better understand how the sentiment of AI-suggested responses affects conversational language, we conducted a second experiment. Using a between-subjects design, we randomly assigned 299 pairs to discuss a policy issue using our app in one of four conditions: with Google smart replies (i.e., participants receive algorithmic responses generated using the Google Reply API \cite{smartReplyAPI}), positive smart replies (i.e., participants receive algorithmic responses that have positive sentiment, as rated by crowdworkers), negative smart replies (i.e., participants receive algorithmic responses that have negative sentiment, as rated by crowdworkers), or no smart replies (i.e., participants do not receive algorithmic responses). We measured conversation sentiment using VADER, a lexicon and rule-based sentiment analysis tool that is ideal for analyzing short, social messages \cite{Hutto2014}. A robustness check with another sentiment analysis dictionary is presented in the Methods section. We aggregated VADER scores into a sentiment polarity score ranking from most positive (1) to most negative (-1), with neutral (0) in the middle. The average conversations comprised 20 messages (\textit{sd}=8.6) and lasted 6.33 minutes (\textit{sd}=2.67).

\begin{figure}[h!]
		\centering
		\includegraphics[width=.5\linewidth]{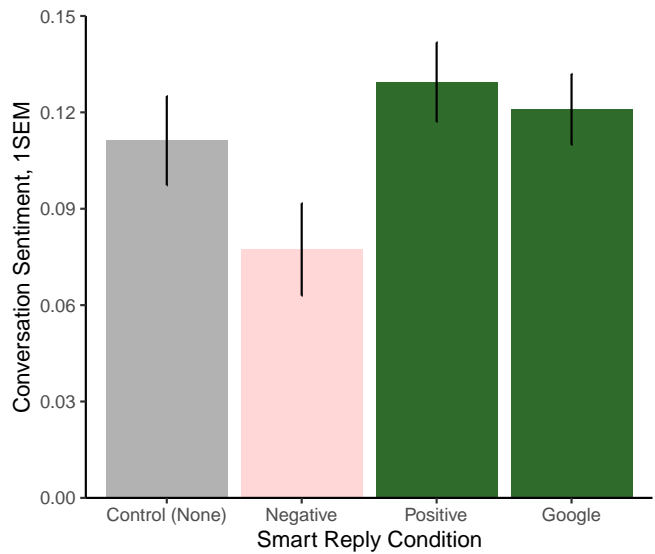}
		\caption{Mean overall conversation sentiment by experimental condition: both participants assigned to no smart replies, negative, positive, or Google smart replies. Error bars show 1 cluster-robust standard error.}
		\label{fig:study2Bar}
	\end{figure}

We found that the presence of positive and Google smart replies caused conversations to have more positive emotional content than conversations with negative or no smart replies (\textit{t}(127)=2.75, \textit{p}=0.007, \textit{d}=.352; Figure~\ref{fig:study2Bar}). Moreover, the finding that widely-used Google smart replies have a similar effect on conversation sentiment as a set of positive (\textit{t}(150)=0.51, \textit{p}=0.61) but not negative smart replies (\textit{t}(127)=2.40, \textit{p}=0.018) highlights the positive sentiment bias of smart replies in commercial messaging apps. Taken together, these findings demonstrate how AI-generated sentiment affects the emotional language used in human conversation. 

\section*{Discussion}
%Communication is the basic means through which we form perceptions of others \cite{mairesse2006automatic, mairesse2006words, mairesse2007using}, build and maintain social relationships \cite{pennebaker2003psychological}, and achieve cooperative outcomes \cite{Zhang2018}. 
Our research shows that a commercially-deployed AI can fundamentally reshape how people communicate with others, and this has both positive and negative consequences. We find that people choose to use AI when given the opportunity, and this increases the efficiency of communication and leads to more emotionally positive language. However, we also find that as participants \textit{think} that their partner is using more algorithmic responses, they perceive them as less cooperative and feel less affiliation towards them. This finding could be related to common assumptions about the negative implications of AI in social interactions. For example, humans are already predisposed to trust other humans over computers \cite{Promberger2006}, and most current communication systems featuring AI mediation lack transparency for users (i.e., the sender knows that their responses have been modified or generated by AI, while the receiver does not). Taken together with users' preference for reducing uncertainty in interactions \cite{Berger1975}, this could lead to negative perceptions of AI in everyday communication. Indeed, these negative perceptions are confirmed by recent work, such as \cite{Hohenstein2018}, where users described how smart replies often did not capture what they wanted to say and could be altering the way that they communicated with others, and \cite{Jakesch}, where text \textit{suspected} of or labeled as being written by AI was perceived as less trustworthy.

Despite negative perceptions about AI in communication, we find  that as people \textit{actually} use more algorithmic responses, their communication partner has more positive attitudes about them. Even though \textit{perceived} smart reply use is viewed negatively, \textit{actual} smart reply use results in communicators being viewed as being more cooperative and affiliative. In other words, it seems that the negative perception of using AI to help us communicate do not match the reality.

%Further work is needed to determine the ways in which developers can improve perceptions of AI in human communication to match its actual effects. One possible avenue for future exploration could involve transparency and "explainability". As previously discussed, users of messaging apps that feature smart replies currently get no information about how these messages are generated, which could be increasing uncertainty and negatively affecting perceptions of smart reply use (e.g., \cite{Shin2019}). If users are instead given background information or real-time feedback on how AI suggestions are being generated, this could potentially alleviate some of the negative perceptions of smart reply use that we have seen. However, as noted in \cite{hancock2020ai}, the amount and type of information that needs to be disclosed for such a system to be considered sufficiently transparent warrants further investigation.

Our work provides evidence that AI can alter the language that people use when interacting with others. Understanding the impact on language is important because language is inextricably linked with listeners' characterizations of a communicator, including their personality \cite{mairesse2006automatic, mairesse2006words, mairesse2007using}, emotions \cite{liscombe2003classifying, pierre2003production}, sentiment \cite{breck2007identifying, pang2005seeing, popescu2005opine, turney2002thumbs}, and level of dominance (i.e., expressing more aggressive instead of affiliative behavior in an interaction) \cite{rienks2005dominance}.  Indeed, we find that AI-generated responses changed the expression of emotion in human conversations. The influence of AI on human emotional communication is deeply concerning given that AI already writes about 6.7 billion emails on our behalf daily \cite{kraus2018mashable}. With the increasing popularity of other forms of AI mediating our everyday communication (e.g., Smart Compose \cite{smartCompose}), we have little insight into how regularly people are allowing AI to help them communicate or the potential long-term implications of the interference of AI in human communication. Our work suggests that interpersonal relationships are likely to be affected, potentially positively. However, the demonstrated changes in language suggest that we could potentially lose our personal communication styles, with language becoming increasingly homogeneous over time. While current implementations of AI in messaging apps increase efficiency by allowing users to respond to messages more quickly, smart reply use is still viewed negatively, and as we have now demonstrated, it has the ability to alter our language when communicating with other humans. 

Our work has implications for the development of AI systems and highlights both opportunities and risks of deploying such systems. A recent laboratory study has shown that a humanoid robot can improve interpersonal communication when expressing vulnerability within a team \cite{traeger2020vulnerable}. Our work takes this research further by demonstrating how a commercially-deployed AI can influence interactions in positive ways through much more subtle forms of intervention than a robot's overt behavior. Merely providing suggestions changes the language used in a conversation, and the changes are consistent with the linguistic qualities of the algorithmic responses. Additionally, previous work has shown that when conversations go awry, people trust the AI more than the person that they’re communicating with and assign some of the blame that they otherwise would have assigned to this person to the AI \cite{hohenstein2020ai}. Taken together, these findings suggest possible opportunities for developers to affect conversational dynamics and outcomes by carefully controlling the linguistics of smart replies that are shown to users,  such as in \cite{Sukumaran2011}. On the other hand, the finding that changes in language are consistent with changes in smart replies raises potential risks as AI gains continues to gain influence over the our social interactions. Knowing that AI can shape the way that we communicate, it is important for researchers and practitioners to consider the broader social consequences when designing algorithms that support communication.

Overall, while AI has the potential to help people communicate more efficiently and improve interpersonal perceptions in everyday conversation, users should be cautioned that these benefits are coupled with alterations to the emotional aspects of our language and a corresponding potential loss of personal expression. 

\section*{Methods}
\label{methods}

The study procedure and all materials were approved by our Institutional Review Board (1610006732), and the study was pre-registered on AsPredicted (40389). 

\subsection*{Web-Based AI-MC Platform}
\label{webBasedAIMCplatform}

	\begin{figure}[h!]
		\centering
		\includegraphics[width=.6\linewidth]{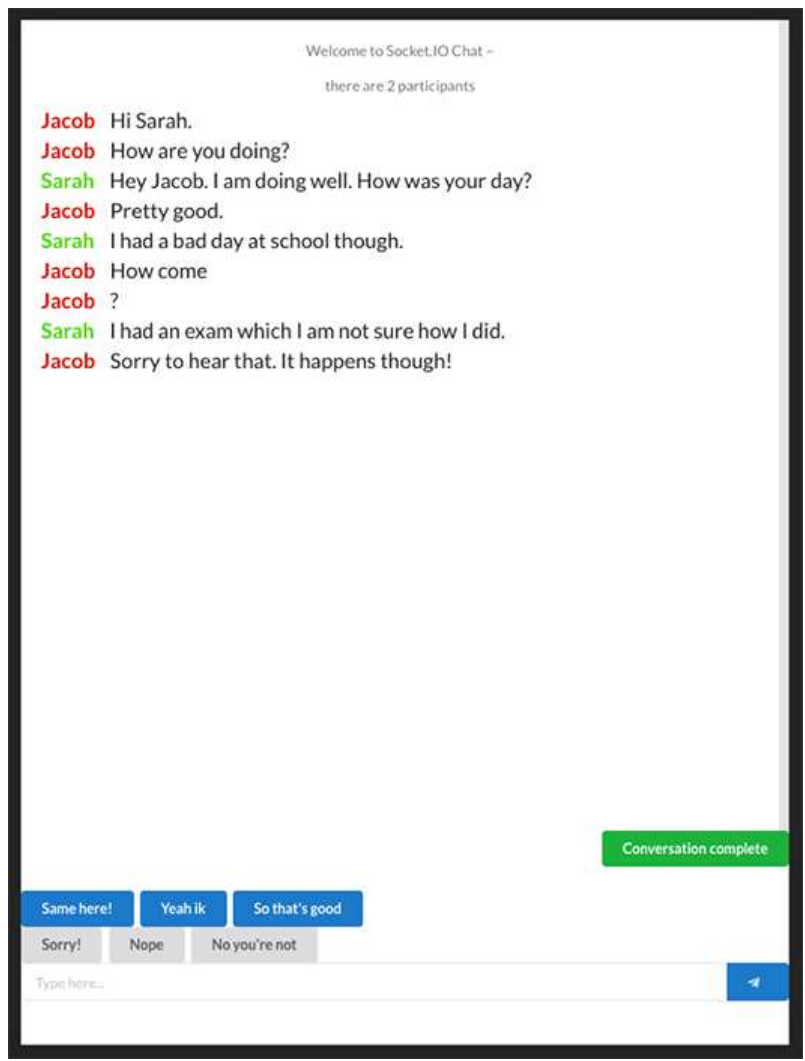}
		\caption{We can use our web-based AI-MC platform to control and record the smart replies that participants see. This figure shows both positive and negative sentiment smart reply examples (i.e., blue and grey boxes, respectively). During actual use, participants see only one of those sets.}
		\label{fig:webApp}
	\end{figure}
	
To develop a nuanced and systematic understanding of the mechanisms by which smart replies and their linguistic properties as well as presentation modalities affect communication, we developed a flexible web-based research platform that allows us to recruit participants online (e.g., through crowdsourcing platforms) and engage them in real-time interpersonal communication tasks while receiving smart reply support, as shown in Figure \ref{fig:webApp}.

The platform is designed as a web application that allows two participants to text chat with one another in real time and runs on all major modern browsers (e.g., Google Chrome 60, Mozilla Firefox 54, Microsoft Edge 14 and Apple Safari 10). It is built using Node.js and MongoDB on the backend with jQuery and Semantic UI framework on the client side. The interface is reactive to device type and resizes itself to work well on desktops, tablets, and mobile devices (Android and iOS). Throughout the design process, we elicited feedback from colleagues to ensure that the application seemed natural and easy to use. Like in existing commercial messaging applications that feature smart replies, in addition to the standard text box for typing messages, participants can also receive smart replies that they can tap to send. Participants can also scroll to see the history of their conversation at any point during the chat. 

Four implementations of the messenger were used in this work: positive and negative sentiment smart replies, real smart replies (i.e., generated by the Google Reply API \cite{smartReplyAPI}), and no smart replies. 

In the positive and negative sentiment smart replies conditions, the smart replies shown to participants had only positive or negative sentiment, respectively. For example, in the positive condition, a participant might see smart replies such as, “I like it” and “I can’t agree more”, whereas in the negative condition, a participant might see smart replies such as, “I don’t get it” and  “No you are not”. These smart replies are chosen randomly from an input json file without being too repetitive (i.e., all three utterances shown in each instance are different, and the same utterance is not shown in immediately subsequent instances). These utterances were pulled from previous work \cite{Hohenstein2018} where crowdworkers rated the sentiment of smart replies, and the suggestions included only those that were rated as having definitive positive or negative sentiment, respectively. In both of these implementations, each time participants sends or receives a message, the smart replies are updated.

In the implementation that did not include smart replies, which served as our control condition, participants had to manually type each message that they sent. 

The final implementation uses Google's Reply model \cite{smartReplyAPI} to generate smart replies. However, since this model does the pre- and post-processing tasks during run-time and its framework is built with C++ and compiled into an Android archive, it is not possible to run it on desktop environments. A stand-alone CPython library on the top of the Reply model can be compiled on a Linux operating system \cite{smartReplyRepo}, which we used to generate smart replies using Google’s Reply model. When users send or receive a message, the Python API receives that message and generates smart replies through the Reply model.

\subsection*{Study 1}
\subsubsection*{Participants \& Procedure}
We collected data from Mechanical Turk participants (\textit{N}=438, 33.7\%F, 0.005\% non-binary) who received monetary payment for their participation. Participants ranged in age from 18-68 (\textit{M}= 34.15, \textit{SD}=10.1). 

The survey itself was conducted using Qualtrics. After obtaining consent, participants were informed that they would be using a messaging system to complete a task with an anonymous partner. Participants were then presented with a task involving a discussion of unfair rejection of work, an issue that is relevant to all crowdworkers on Mechanical Turk \cite{mcinnis2016taking}. Specifically, we asked pairs to come to an agreement on the ”top 3 changes that Mechanical Turk could make to better handle unfairly rejected work”. Participants were asked to open the web-based messaging platform in another window while still viewing the Qualtrics survey. After opening the messaging platform, participants waited up to 5 minutes for another participant to enter the conversation. If 5 minutes elapsed without another participant arriving, participants were able to prematurely exit the survey and receive partial compensation. Once another participant arrived, the pair was as much
time as they needed to come to an agreement on a ranked list. When finished with the task, participants could press a ”Conversation complete” button in the messenger and receive a conversation completion code that they pasted into the Qualtrics survey to confirm that they had completed a conversation with a partner. 

After verifying that a conversation was completed and giving a brief description of smart replies, we asked participants how much they believed their partner had used smart replies. Participants were also asked to fill out the Perceived Cooperative Communication scale and the Interpersonal Adjective Scale, Revised (IAS-R).
	
Perceived cooperative communication was measured through a 7-item scale \cite{lee1997leader} where participants rated their agreement with statements describing cooperative communication in their overall interaction with their partner. The instructions read, "Thinking about your interaction with your partner, please rate the extent to which you agree with each of these statements". Participants rated each statement on rating-scale items anchored by "Strongly disagree" (1) and "Strongly agree" (7).
	
The IAS-R provides an empirical measure of various dimensions that underlie interpersonal transactions \cite{wiggins1988psychometric}. To shorten the measure, two adjectives with the highest loading factors from each interpersonal octant were selected, based on the analysis of Wiggins et al \cite{wiggins1988psychometric}, resulting in 16 items to be ranked. The instructions read, "Below are a list of words that describe how people interact with others. Based on your intuition, please rate how accurately each word describes your conversation partner" (adapted from \cite{knutson1996facial}). Participants rated each statement on rating-scale items anchored by "Extremely inaccurate" (1), "Somewhat accurate" (4), and "Extremely accurate" (7). These ratings were then combined according to a formula adapted from \cite{wiggins1988psychometric} to determine ratings of affiliation and dominance \cite{knutson1996facial}.
  
The presentation of the 3 post-task measures was randomized between participants to avoid any possible order bias. Lastly, participants were asked about demographic information as well as for any comments that they had about the survey.

\subsubsection*{Statistics}
We analyzed the data using instrumental variable regression with self smart reply use and partner smart reply use instrumented by condition assignment; no covariates were added. We used the \textit{ivreg} function in the R AER package \cite{kleiber2020package}. We computed cluster-robust standard errors (i.e., CR2) using the \textit{coef\_test} function in the R clubSandwich package \cite{pustejovsky2017clubsandwich}. The reported and plotted estimates represent coefficients, t-statistics and p-values from the IV regression output.

\subsubsection*{Analysis and Results}
\label{analysisResults}
We excluded conversations from the analysis that had less than 10 messages exchanged overall and where one participant sent less than 3 messages. For the analysis of post-conversation self-report outcomes, we also excluded participants who did not complete the full survey.

Sentiment was analyzed using VADER, a lexicon and rule-based sentiment analysis tool specifically attuned to sentiments expressed on social media \cite{Hutto2014}. This analysis tool yields a sentiment metric indicating how positive, negative, or neutral the sentiment of the supplied text is. For our purposes, messages were analyzed individually using the VADER compound sentiment output, an aggregated score ranging from -1 to 1 (i.e., most negative to most positive) based on the three aforementioned sentiment components. 

The formulae for deriving aggregate measures of dominance and affiliation from the IAS-R (adapted from \cite{knutson1996facial, wiggins1988psychometric}) are given below:

1. Octant scores are computed from adjective ratings:

$\>$$\>$$\>$$\>$PA = (dominant + assertive)/2

$\>$$\>$$\>$$\>$BC = (sly + cunning)/2

$\>$$\>$$\>$$\>$DE = (unsympathetic + warmthless)/2

$\>$$\>$$\>$$\>$FG = (unsociable + antisocial)/2

$\>$$\>$$\>$$\>$H1 = (shy + unaggressive)/2

$\>$$\>$$\>$$\>$JK = (uncunning + unsly)/2

$\>$$\>$$\>$$\>$LM = (gentle + tender)/2

$\>$$\>$$\>$$\>$NO = (friendly + outgoing)/2

2. Dominance and affiliation scores are computed from these octant
scores:

$\>$$\>$$\>$$\>$DOM = PA - HI + .707(NO + BC - FG - JK)

$\>$$\>$$\>$$\>$AFF = LM - DE + .707(NO - BC - FG + JK)

\subsection*{Study 2}

\subsubsection*{Participants \& Procedure}
\label{study1ParticipantsProcedure}
We collected messaging conversations from participants (\textit{N}=599, 37.2\% F, 0.1\% non-binary) through Mechanical Turk who received monetary payment for their participation. Participants ranged in age from 19–69 (\textit{M}=35.6, \textit{SD}=9.96). To ensure that any language differences that we found were not the result of demographic differences between the four conditions \cite{schwartz2013personality}, we examined the demographic makeup (i.e., age, gender, and race) between conditions and did not find any significant differences.

The survey and procedure were similar to the previous study, except participants in the AI-mediated messaging condition were  informed that they would be ”[...] using an AI-mediated messaging system to have a conversation with your partner. While you are messaging, artificial intelligence (AI) will provide smart replies that you can simply tap to send.”, while participants in the control condition were told that they would be ”[...] using a standard messaging system to have a conversation with your partner.” After verifying that a conversation was completed, participants were asked about demographic information as well as for any comments that they had about the survey.

\subsubsection*{Statistics}
We analyzed the resulting data at the individual level using a simple linear regression with cluster-robust standard errors using the \textit{lm\_robust} function in the R estimatr package \cite{blair2018package}. The dependent variable was the individual language measure (i.e., VADER sentiment) and the independent variable was the assigned condition; no covariates were added. The reported statistics are the t-statistic and p-value for the relevant coefficient, and Cohen’s d computed manually.

\subsubsection*{Robustness Check for Sentiment Results}
We presented an analysis of conversation sentiment using VADER, a lexicon and rule-based sentiment analysis tool that is ideal for analyzing short, social messages \cite{Hutto2014}. As in Study 1, we excluded conversations from the analysis that had less than 10 messages exchanged overall and where one participant sent less than 3 messages. 

To ensure that results do not significantly change with other dictionaries, we performed a robustness check using Linguistic Inquiry
and Word Count (LIWC), a dictionary-based text analysis tool that determines
the percentage of words that reflect a number of linguistic processes,
psychological processes, and personal concerns \cite{Hutto2014}. To verify our findings with respect to sentiment from VADER, we analyzed Affect scores from LIWC. Affect, with values ranging from 0-100, is made up of
Positive and Negative Emotion variables, which also range from 0-100. For example,
a message with an Affect score of 50 could be made up of a Positive
Emotion score of 50, a Positive Emotion score of 25 and a Negative Emotion
score of 25, or a Negative Emotion score of 50.

	\begin{figure}[h!]
		\centering
		\includegraphics[width=.5\linewidth]{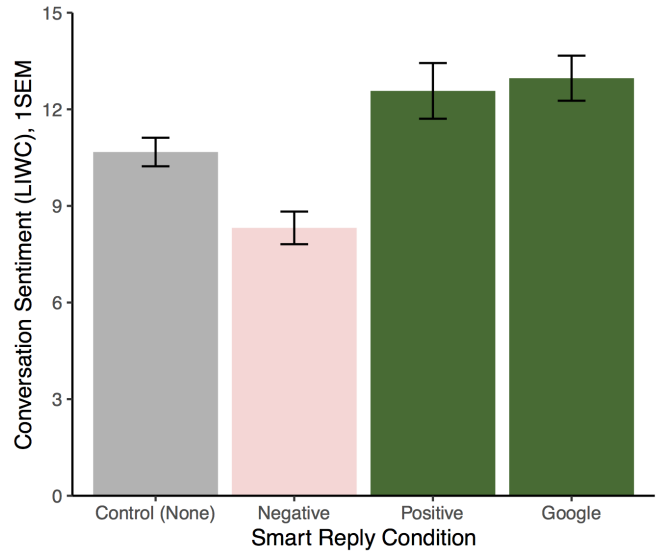}
		\caption{Mean overall conversation affect by experimental condition: both participants assigned to no smart replies, negative, positive, or Google smart replies. Error bars show 1 cluster-robust standard error.}
		\label{fig:LIWC}
	\end{figure}

All findings with respect to VADER sentiment were confirmed using LIWC. We found that the presence of positive and Google smart replies caused conversations to have higher affect than conversations without smart replies (\textit{t}(124)=2.95, \textit{p}<0.001, \textit{d}=0.272). The effect of positive and Google smart replies on affect was statistically similar (\textit{t}(150)=0.354, \textit{p}=0.724). The presence of negative smart replies had a strong negative effect on conversation affect compared to the control condition without smart replies (\textit{t}(123)=-3.50, \textit{p}<0.001, \textit{d}=0.454).

\subsection*{Limitations}
There were several limitations to this work. First, we analyzed conversations from participants completing a contrived task on Mechanical Turk. Although we attempted to choose a task that would be personally relevant to all crowdworkers and effectuate the interpersonal closeness that we hoped to examine, many other types of everyday messaging conversations exist, and future work should examine how these results hold up in disparate contexts.

Since our web-based messenger is not yet robust enough for mobile use, this work focused specifically on AI-mediated messaging conversations in a desktop computer environment and may not generalize to similar messaging situations on other use contexts. Interpersonal perceptions in mobile messaging contexts featuring smart replies should also be examined.

As is standard in similar literature (e.g., \cite{wiggins1988psychometric}), interpersonal perceptions were measured as momentary states. However, these perceptions change and develop over time, so future work should examine whether and how these measures are affected longitudinally under the influence of smart replies. Similarly, these studies occurred with anonymous crowdworkers completing a one-time interaction. We do not know whether our findings would be different in relationships with various levels of interpersonal closeness. Future work should investigate how interpersonal perceptions are related to smart reply use in more socially intimate relationships, such as between friends or co-workers. Additionally, we investigated interpersonal perceptions resulting from real-time messaging conversations, which could manifest differently in other communication contexts. Future work should examine how interpersonal relationships are affected by the presence of AI mediation in asynchronous communication contexts, such as email.

\section*{Data Availability}
The data has been deposited in the Mendeley Data repository (DOI: 10.17632/6v5r6jmd3y.1 \cite{mendeleyData}).

\bibliography{scibib}

\section*{Acknowledgements}

We would like to acknowledge Hirokazu Shirado and Michael Macy for providing feedback on this work and support from grants from the National Science Foundation (Award Numbers IIS-1901151 and 72517).

\section*{Author contributions statement}

J.H., D.D., R.K., and M.J. conceptualized the experiments, J.H., D.D., R.K., Z.A., and M.J. curated the data, J.H. and R.K. performed the formal analysis, J.H. performed the investigation, J.H., R.K., and M.J. determined the methodology, J.H. and M.J. administered the project, J.H., D.D., and Z.A. provisioned resources, J.H., D.D., and Z.A. developed software, J.H. and M.J. wrote the original draft, K.L., M.N., J.H., and M.J. acquired funding, and all authors reviewed and edited the manuscript.

\section*{Additional information}

\textbf{Competing interests}: Authors declare no competing interests.

\end{document}